\documentstyle[12pt]{article}
\begin{document}
\baselineskip 0.5cm
\parskip 0.3cm
\parindent=0pt
\parskip 0.3cm
\parindent 0pt
\def\ref{\hangindent=1cm \hangafter=1 }
\vskip 0.3cm
{\bf\centerline{The optical counterpart to  $\gamma$-ray burst GRB970228}}
{\bf\centerline{observed using the Hubble Space Telescope}}
\vskip 1cm
{\centerline{Kailash C. Sahu$^1$, Mario Livio$^1$, Larry Petro$^1$, 
F. Duccio Macchetto$^1$,}}
{\centerline{Jan van Paradijs$^{2,5}$,  Chryssa Kouveliotou$^3$, 
Gerald J. Fishman$^4$,  
}}
{\centerline{Charles A. Meegan$^4$, Paul J. Groot$^2$, Titus Galama$^2$ }}

{{$^1$Space Telescope Science Institute,}
{3700 San Martin Drive, Baltimore, MD 21218, USA}}

$^2$Astronomical Institute ``Anton Pannekoek", University
of Amsterdam, \& Center for High Energy Astrophysics, Kruislaan 403,
1098 SJ Amsterdam, The Netherlands.

$^3$Universities Space Research Association, NASA Marshall Space 
Flight Center,  ES-84, Huntsville, AL 35812, USA 

$^4$NASA Marshall Space Flight Center, ES-81, Huntsville, AL 35812, USA 
 
$^5$Physics Department, University of Alabama in Huntsville,
Huntsville, AL 35899, USA.
\vskip 1cm
{\bf{\centerline{To appear in Nature (29 May 1997 issue)}}}
\smallskip
{\centerline{(Submitted: March 28, 1997; Revised: April 9, 1997; Accepted: April 24, 1997)}} 
\vfill \eject
{\bf Although more than 2,000  astronomical gamma-ray bursts (GRBs) 
have been  detected, and numerous models proposed to explain their 
occurrence$^1$, they have remained enigmatic owing to the lack of an
obvious counterpart at other wavelengths$^{2-5}$.
The recent ground-based detection$^{6,7}$ 
of a transient source in the vicinity of 
GRB 970228$^{8-11}$ may therefore have provided a breakthrough.
The optical counterpart appears to be embedded in an extended source which,
if a galaxy as has been suggested$^{7,12}$,
would lend weight to those models that place GRBs at cosmological
distances. 
Here we report the observations using the Hubble Space Telescope 
of the transient counterpart and extended source
26 and 39 days after the initial $\gamma$-ray outburst.
We find that the counterpart has faded since the initial detection
(and continues to fade), but the extended source exhibits
no significant change in brightness between the two dates of
observations reported here.
The size and apparent constancy
between the two epochs of HST observations 
imply that it is extragalactic, but its 
faintness makes a definitive statement about its nature difficult.
Nevertheless, the decay profile of the transient source is consistent
with a popular impulsive-fireball model$^{13}$,
which assumes a merger between two neutron stars in a distant galaxy.}

The optical counterpart to GRB 970228 was observed with the HST Wide Field and
Planetary Camera (WFPC2), on March 26$^{14}$ and April 7$^{15}$, about 26 and 
39 days after the outburst, respectively. 
The optical counterpart  was placed at the centre of the
Planetary Camera (PC) field of view, so as to attain maximum spatial
resolution. Three other candidates (one radio$^{16}$ and two
optical$^{17-22}$) proposed earlier as counterparts of GRB 970228 are
also within the WFPC2 field of view. Observations were taken in the
F606W (wide V) and F814W (I) filters$^{23}$. Four exposures with a
total integration time of 4700 seconds were taken in the F606W filter
and two exposures with a total integration time of 2400 seconds were
taken in the F814W filter, during each observing run. The images were
corrected for bias and flat-field variations through the standard
Space Telescope processing software. The number of images was sufficiently large to
allow for a proper cosmic ray rejection from the flat- fielded images. The
images were combined after cosmic ray rejection. Fig. 1a shows a part
of the combined F606W and F814W images (for the two observations)
where the optical counterpart of the GRB can be seen at the centre.
The optical counterpart is embedded in an extended source which can be
seen in both passbands.  An examination of the PSF of different
sources also clearly shows the extended nature of this
source. As seen in Fig. 1a, the angular extent of the extended
source is $\sim 1$ arcsecond and it is elongated in the $E-W$
direction.  The point source lies about 0.3 arcseconds south of the
centre of the extended source.  Photometry in both wavebands was
carried out for the point source using the photometric software
package DAOPHOT. The derived magnitudes for the point source are given
in Table 1.  As noted earlier, this source was first detected 
in the optical wavelengths about 1
day after the outburst, when the magnitudes were $V = 21.3$ and $I =
20.6$ and the magnitudes observed 9 days after the outburst were $V >
23.6$ and $I > 22.2$. The observed decline in brightness is shown in
Fig. 2 with the predictions of various models (see discussion below).
The $V-I$ colour of the point source has increased by about 0.6
magnitude in 25 days, which is consistent with a cooling trend, and
then remained roughly constant for the following 12 days. 

We tested for a possible proper motion of the point source between the
two epochs (March 26 and April 7) in the two filters.
We find a potential motion, both in the V and I filters,
 in the NW direction of about 0.006 arcseconds. Given the errors involved
(this represents a 1.5 to 2 $\sigma$ result), 
and considering the fact that the GRB is the faintest object
in the field and is embedded in a nebulosity, we conclude that
no proper motion has been detected. 

We would like to point out that even without discussing the
probability of finding a variable star in the error box of the GRB
during the given time frame, it is possible to rule out the
possibility that the optical transient is actually an unrelated nova,
dwarf nova, flare star, or a supernova. The evolution of the colours is
inconsistent with that of novae (the transient becoming redder rather
than bluer following maximum)$^{24}$, and the value of (V-I) (which is
$\sim$0.0 for dwarf novae), and its development are inconsistent with
those of dwarf novae$^{25}$  (correction for absorption indicated by
the LECS  (0.1 - 10 kev) observations on board the
SAX satellite does not change this conclusion). The duration of the
event is longer by orders of magnitude than those of stellar
flares$^{26}$. Although supernovae decay in 
visual wavelengths at a slower rate than observed for the optical
transient, they can decay in the UV as fast as 2.5 magnitudes per
day$^{27}$. Thus, a supernova at the appropriate redshift (z $\sim$
1--2), could appear to decay in the visual (UV rest frame) at the
observed rate. However, the observed fast decline in I would imply
that the redshift of this supernova is $>1$. Such a large redshift
would make this object brighter than the brightest supernovae
(observed either at low$^{28}$, or at high$^{29,30}$ redshift) by
about 2 magnitudes at the peak, thus making it improbable to be a
supernova. 

In order to determine the magnitudes for the extended source, the
point source was subtracted using a PSF derived from other point
sources in the images.  DAOPHOT was used to derive the magnitudes of
the extended source within a radius of $0.6$ arcseconds.  The
resultant magnitudes are given in Table 1.  The angular dimensions of
the extended source are consistent with earlier estimates from the
ground; however, its brightness appears lower by about 0.9 magnitudes
in the March 26th observation.  At present it is not entirely clear
whether this discrepancy represents a real decline, or whether it
simply results from the inability of the ground-based observations to
resolve the point and extended sources.  In order to investigate this
further, we performed an experiment to simulate the ground-based
observations of March 9.  We used the combined F606W image and
artificially increased the brightness of the point source to $V =
24.0$ (as predicted by the best theoretical models, see below).  We
then smoothed the image to simulate 1-arcsecond seeing from the
ground.  In this simulated image the extended nature of the source is
barely discernible and the brightnesses of the individual components
cannot be separately determined. We should note that visual 
comparison of the images of the extended source taken on 
March 26 and April 7 reveals small differences. Taken at face value,
these differences could be interpreted as indicating changes
in either the relative brightness of different parts of the
extended source, or in its position. However, no such
definitive statement can be made, given the errors involved
in the fluxes. This conclusion is further strengthened 
 by the following exercise. We subtracted the image of April 7 
from that of March 26, and the result is presented in Fig. 1b.
As can be seen, the point source is clearly visible in the subtracted image
(which is consistent with its decline in brightness), while no trace of the
extended source may be distinguished from the background noise.  
Thus, within the errors, it appears that the extended source 
(which dominates the light after March 13) remained constant.

\begin{table}
\begin{flushleft}
{\caption[]{HST Photometry of GRB 970228 Field$^*$}}
\begin{tabular}{llccccccl}
\hline
\\
Source  & Date (UT) & V & I&\\
\hline
\\
Point Source  &Mar. 26.4 & 26.1 $\pm${0.1}&  24.2 $\pm${0.1}&\\
Point Source  &Apr.  7.2  &26.4 $\pm${0.1}  &24.6 $\pm${0.1}&\\
Ext. Source   &Mar. 26.4  &24.9 $\pm${0.3}  &24.5 $\pm${0.3}&\\
Ext. Source   &Apr.  7.2  &25.2 $\pm${0.35}  &24.3 $\pm${0.35}&\\
\hline
\end{tabular}
\end{flushleft}

$^*$ The magnitudes given here are in the Cousins bands,
where the magnitudes have been transformed from WFPC2
to Cousins filters taking an appropriate colour correction into account.
For colour correction, an M2 type spectrum is assumed for the
point source and an F2 type spectrum is assumed for the 
extended source. Since Cousins I filter is similar to F814W,
the color correction is close to zero in the I band. 

\end{table}

\medskip
If the extended source was indeed declining (or changing in position), 
the implication would 
have been that it is probably powered by the GRB.  Since the angular
dimensions of the source are about 1 arcsecond, a brightness variation
in 15 days would imply a distance to the source of no more than about
2.7 kpc (assuming that the optical light comes from 
the source location, and is not light scattered from an intervening 
``screen"). Such a short distance would not be consistent even with models
which place GRBs in an extended Galactic halo$^{31}$, since these
models require typically distances significantly larger than the distance
between Earth and the Galactic centre.  Models placing all the
GRBs in the
Galactic disk have been convincingly ruled out by observations$^{32}$.
Clearly, with only one optical counterpart to date, we cannot rule out
the possibility of the existence of two populations of GRBs$^{33,34}$,
one at cosmological distances and the other of Galactic origin.  The
above discussion does suggest that the extended source
remained constant and that the apparent decline in brightness is
merely a consequence of the low resolution of the ground-based observations.
 If this interpretation is correct, then the extended source could be
an external galaxy (its colour is inconsistent with it being
Galactic cirrus as seen in the infrared observations of IRAS). We should also note that the observations of 
SAX LECS indicate an absorption towards the source of A$_v$ $\simeq$
2.5 magnitudes. With the foreground extinction 
being$^{35}$ A$_v$ = 0.4 $\pm$ 0.3,
this could indicate absorption in a galaxy (although the
possibility of circumstellar absorption in the vicinity of
the source cannot be ruled out).  

Since we have found that the optical counterpart may be
associated with a galaxy, we first have to determine the probability
of a chance superposition. The latter can be estimated by the
fractional area covered by galaxies of 25th  magnitude (in V), or brighter.
From the catalog computed from the Hubble Deep Field (HDF)$^{36}$
using the software package FOCAS, we derive a probability for chance superposition of the order of 
4\%. This is consistent with the probability estimates of van 
Paradijs et al.$^{7}$

In the following we will therefore assume that the GRB is located in
an external galaxy and we will examine the implications of this assumption
for theoretical models of GRBs. 

One of the key components that can lead to a better understanding of
GRBs is the question of their location.  The isotropy and
inhomogeneity in the distribution of the bursts have led to two
leading classes of models, one placing the sources at cosmological
distances$^{32}$, and the other placing them in the extended Galactic
halo$^{31}$.  The association of GRB 970228 with an external galaxy
would clearly support the cosmological models. 

A second question that can be addressed is whether GRBs are associated
with nuclear activity of active galactic nuclei (e.g. the burst being
produced by tidal disruption of a star$^{37}$). The HST observations
clearly show that the optical counterpart is not located at the centre
of the brightness distribution, suggesting that GRB 970228 is not at
the centre of the host ``galaxy." Although inconclusive, this suggests
that GRBs as a class are not related to central supermassive black holes. 

Irrespective of the nature of the extended source
(but assuming that the optical transient point source is indeed the
GRB), the data provide
valuable information on the evolution of GRB remnants. In particular,
specific predictions for the time behaviour of the optical emission
from cooling and expanding fireball ejecta (in the context of cosmological 
models) were made by Meszaros and
Rees$^{13}$ and (for X-rays in the context of Galactic halo models) by 
Liang et al.$^{38}$. In Fig. 2, we show the
predictions of the impulsive fireball models of Meszaros and Rees and
the model by Liang et al., in which cooling of the non-thermal leptons
occurs by saturated Compton upscattering, together with the optical
observations. As seen in the figure,  our results are
consistent with a simple impulsive model$^{13}$ in which only the forward
blast wave radiates efficiently and in which the peak frequency drops
below the optical band about 1.7 days after the burst (t$_{opt} \sim
1.7$ days). It is
important to note that the observations rule out models$^{13}$ in
which the energy input is continuous rather than impulsive, in their
simplest form. These
models produce either no optical flux at all after the gamma-ray
flash, or a flux which declines like t$^{-6}$, much faster than
observed. We should also note that the blast wave models predict a
change in their power law (towards a steeper decline) after the blast wave
has snowplowed through a rest-mass energy of the order of the burst energy.
This predicts a change in the power law after a few days, for a burst energy of
10$^{42}$ ergs (corresponding to extended Galactic halo models). 
No such change has been observed. If anything, a change in the power law to
a shallower decline may have been detected$^{39}$ on March 6. This tends to
support a cosmological origin for the burst (in which case a change is
expected only after a time scale of years). 

The fact that cosmological models require peak luminosities of the
order of 10$^{51}$ erg s$^{-1}$ has led to the popularity of models
involving the mergers of either two neutron stars, or a neutron star
and a black hole$^{40,41}$. The frequencies of such mergers have been
estimated by Phinney$^{42}$ and by Narayan, Piran and Shemi$^{43}$.
Using the ``best guess" values for the parameters from Phinney and a
Hubble constant of H$_0$ = 65 km s$^{-1}$ Mpc$^{-1}$, we obtain for
the ratio of the neutron star merger rate in disk galaxies to that in
ellipticals R$_{disk}$/R$_{elliptical} \simeq 80$. Thus, if GRBs are
produced by such mergers, then binaries in disks should dominate the
observed rate. A close examination of the ``host galaxy" of GRB 970228
(Fig. 1a) reveals indeed elongated features to the east, which are
suggestive of a spiral, or irregular morphology, rather than an
elliptical one.  Furthermore, the $V - I$ colour of the extended
source (admittedly uncertain) suggests a late-type galaxy. 

A definitive answer to the nature of the extended source associated 
with GRB 970228 will be provided by HST observations after a longer 
time interval.

\vskip 1cm
{\it Acknowledgements:} We would like to thank Bob Williams for the allocation of 
Director's Discretionary time for the observations, Mike Vogeley for help 
with the HDF, and Andy Fruchter, Massimo Stiavelli, Stefano Casertano 
and Ron Gilliland and Ralph Wijers for helpful discussions. We acknowledge the help of 
Inge Heyer, Matt McMaster, Mike Potter and Zolt Levay in the reduction of the data.  
We would like to thank Luigi Piro and Enrico Costa for alerting us about 
the discovery of the GRB source.

\medskip
{\parskip 0pt
{\bf References:}
\vskip 0.1cm

\ref 1. Nemiroff, R.J. A Century of Gamma Ray Burst Models. Comments
on Astrophys. {\bf 17}, 189-205 (1994). 

\ref 2. Fishman, G.J., \& Meegan, C.A. Gamma-Ray Bursts. Ann. Rev.
Astron. Astrophys. {\bf 33}, 415-458 (1995).\par 

\ref 3. Hurley, K. Gamma-Ray Burst Observations: Past and Future. in 
{\it Gamma Ray Bursts} (Conf. Proc. 265, Am. Inst. Phys. 
New York, eds Paciesas, W. \& Fishman, G.), 3-12  (1992).\par 

\ref 4. Higdon, J. \& Lingenfelter, R. Gamma-Ray Bursts. Ann. Rev.
Astron. Astrophys. {\bf 28}, 401-436 (1990).\par 

\ref 5. Hartman, D. Gamma-Ray Burst Observations: Theoretical 
Considerations. in {\it The Gamma Ray Sky with COMPTON and SIGMA},
NATO ASI Proc., ed. M. Signore, P. Salati, G Verdrenne, 329-367
(1995).\par

\ref 6.  Groot, P.J., et al. IAU Circ. No. 6584 (1997). 

\ref 7. van Paradijs, J., et al. Transient Optical Emission 
from the Error Box of the $\gamma$-Ray Burst of 28 February 1997. 
Nature, {\bf 386}, 686-689 (1997).

\ref 8.  Costa, E., et al. IAU Circ. No. 6572 (1997). 

\ref 9.  Costa, E., et al.  IAU Circ. No. 6576 (1997).  

\ref 10.  Cline, T.L., et al. IAU Circ. No. 6593 (1997). 

\ref 11.  Hurley, K., Costa, E., Feroci, M., Frontera, F., Dal Fiume, D., 
\& Orlandini, M.  IAU Circ. No. 6594 (1997). 

\ref 12. Metzger, M.R., Kulkarni, S.R., Djorgovski, S.G., Gal, R., Steidel, 
C.C., \& Frail, D.A. IAU Circ. No. 6588 (1997). 

\ref 13. Meszaros, P., \& Rees, M. Optical and Long-Wavelength Afterglow from 
Gamma-Ray Bursts. Astrophys. J. {\bf 476}, 232-235 (1997).

\ref 14. Sahu, K.C., Livio, M., Petro, L., \& Macchetto, F.D. IAU Circ. No.
6606 (1997). 

\ref 15. Sahu, K.C., Livio, M., Petro, L., \& Macchetto, F.D. IAU Circ. No.
6619 (1997). 

\ref 16. Frail, D.A., et al. IAU Circ. No. 6576 (1997).

\ref 17.  Groot, P.J., et al.  IAU Circ. No. 6574 (1997).

\ref 18.  Margon, B., Deutsch, E.W., \& Secker, J. IAUC No. 6577 (1997).  

\ref 19. Pederson, H., et al.  IAU Circ. No. 6580 (1997).

\ref 20. Wagner, R.M. \& Buie, M.W. IAU Circ. No. 6581 (1997).

\ref 21. Wagner, R.M., Foltz, C.B., \& Hewett, P.  IAU Circ. No. 6581 (1997).  
 
\ref 22. Metzger, M.R., Kulkarni, S.R., Djorgovski, S.G., Gal, R., Steidel, 
C.C., \& Frail, D.A.  IAU Circ. No. 6582 (1997). 

\ref 23. Biretta, J.A., et al. WFPC2 Instrument Handbook, Version 4.0,
STScI, Baltimore (1996).

\ref 24. Warner, B. in {\it Cataclysmic Variable Stars}, (Cambridge: Cambridge
University Press), 261 (1995).

\ref 25. Warner, B. in {\it Cataclysmic Variable Stars}, (Cambridge: Cambridge
University Press), 148 (1995).

\ref 26. Mirzoyan, L.V. Optical Flares: Observations and 
Interpretations. in {\it Flares and Flashes}, eds. J. Greiner,
H.W. Duerbeck, \& R.E. Gershberg (Berlin: Springer), 47-54 (1995).

\ref 27. Panagia, N. On the Energetics of SN 1987A.
in {\it Supernova 1987A in the Large Magellanic
Cloud}, eds. M. Kafatos and A.G. Michalitsianos, (Cambridge: Cambridge University
Press), 96-105 (1988).

\ref 28. Panagia, N. et al. Coordinated optical, ultraviolet, radio, 
and X-ray observations of Supernova 1979c in M100. Mon. Not. R. Astron.
Soc., {\bf 192}, 861-879 (1980).

\ref 29. Perlmutter, S, et al. A Supernova at z=0.458 and 
Implications for Measuring the Cosmological Deceleration. Astrophys.
J., {\bf 440}, L41-L44 (1995).

\ref 30. Perlmutter, S., et al. Astrophys. J., in press (1997).

\ref 31. Podsiadlowski, P, Rees, M., \& Ruderman, M. Gamma-ray bursts 
and the structure of the Galactic halo.  Mon. Not. R. Aastr. Soc.
 {\bf 273}, 755-771 (1995).

\ref 32. Mao, S., \& Paczynski, B. On the Cosmological Origin of 
Gamma-Ray Bursts. Astrophys. J. {\bf 388}, L45-48 (1992).

\ref 33. Katz, J. Two Populations and Models of Gamma-Ray Bursts. 
Astrophys. J. {\bf 422}, 248-259 (1994). 

\ref 34. Lamb, D.Q. Gamma-Ray Bursts. in {\it Neutron Stars: Theory and Observation}, eds.
J. Ventura and D. Pines (Dordrecht: Kluwer), 545-560 (1991).

\ref 35. Hakkila, J., Myers, J.M., Stidham, B.J., \& Hartman, D.H. 
A Computerized Model of Large-Scale Visual Interstellar Extinction. 
submitted to Astron. J. (1997).

\ref 36.  Williams., R.E. et al. The Hubble Deep Field:  
Observations, Data Reduction, and Galaxy Photometry. Astron. J. {\bf
112}, 1335 (1996). 

\ref 37. Carter, B. Cosmic Gamma-Ray Bursts from Black Hole Tidal 
Disruption of Stars?   Astrophys. J. {\bf 391}, L67-L70 (1992).

\ref 38. Liang, E., Kusunose, M., Smith, I.A., \& Crider, A. 
Physical Model of Gamma-Ray Burst Spectral Evolution. 
Astrophys. J. {\bf 479}, L35 (1997).

\ref 39. Galama, T.J., et al. The Optical Light Curve of GRB 970228. 
submitted to Nature (1997).

\ref 40. Paczy\'nski, B. Gamma-Ray Bursters at Cosmological 
Distances.  Astrophys. J.  {\bf 308}, L43-L46 (1986).

\ref 41. Eichler, D., Livio, M., Piran, T., \& Schramm, D. 
Nucleosynthesis, neutrino bursts, and ${\gamma}$-rays from coalescing 
neutron stars.   Nature, {\bf 340}, 126-128 (1989). 

\ref 42. Phinney, E.S. The Rate of Neutron Star Binary Mergers in the 
Universe:  Minimal Predictions for Gravity Wave Detectors.  Astrophys.
J.  {\bf 380}, L17-L21 (1991). 

\ref 43. Narayan, R., Piran, T. \& Shemi, A. Neutron Star and Black 
Hole Binaries in the Galaxy. Astrophys. J.  {\bf
379}, L17-L21 (1991). 

\ref 44.  Groot, P.J., et al. IAU Circ. No. 6588 (1997).
}
\vfill \eject
{\bf Figure captions}

{\bf Fig. 1a.} The optical counterpart to GRB 970228 was observed with the HST Wide Field and
Planetary Camera (WFPC2), between March 26.33 and 26.49 UT, about 26
days after the outburst$^{14}$, and again between April 7.15 and 7.32
UT, 39 days after the outburst$^{15}$.
Fig a, shows the sum of the  F606W and F814W images (resolution 0.045 
arcseconds/pixel) taken at both epochs
smoothed with a 3x3 spatial filter. 
The small axes indicate the directions of north (arrow) and east, and 
the size of the image is 11.5 x 11.5 arcsec.  The point source is at the centre of this image, and 
it is embedded in a diffuse source, $0.3$ arcseconds to the north.

{\bf Fig. 1b.} The difference image obtained by subtracting the summed
image of April 7 from the summed image of March 26 (amplified
4 times with respect to {\bf a}. The point source
is clearly seen, but there is no trace of the extended source.
This is consistent with fading of the point source and constancy of
the extended source (within errors) between the two observations.

{\bf Fig. 2} Measured brightness of GRB 970228 optical counterpart and
models of gamma-ray burster remnants.  The brightness of the optical
counterpart has been measured by Groot et al. in the 4.2-m William
Herschel Telescope
discovery image$^{6,7}$ (closed circle, labeled Optical Transient).  The 
HST measurements presented in this Letter are indicated by a closed 
square (point source) and open square (adjacent extended source).  The 
measured brightness$^{6,7,12,44}$ of an extended source at the position 
of the transient is indicated by open circles.  The Meszaros and
Rees models$^{13}$ for the brightness of GRB remnants are labeled MR. 
In model a1, only the forward blast wave radiates efficiently;
in model a2, both the forward and the reverse shocks are efficient radiators;
model a3 is similar to a2, but has a different origin for the magnetic field. 
The Liang et al. model$^{38}$ is labeled LKSC. 
${t - t_{GRB}}$ is the time elepsed since the outburst;
$t_{opt}$ is the time at which the peak frequency drops below the
optical band. 

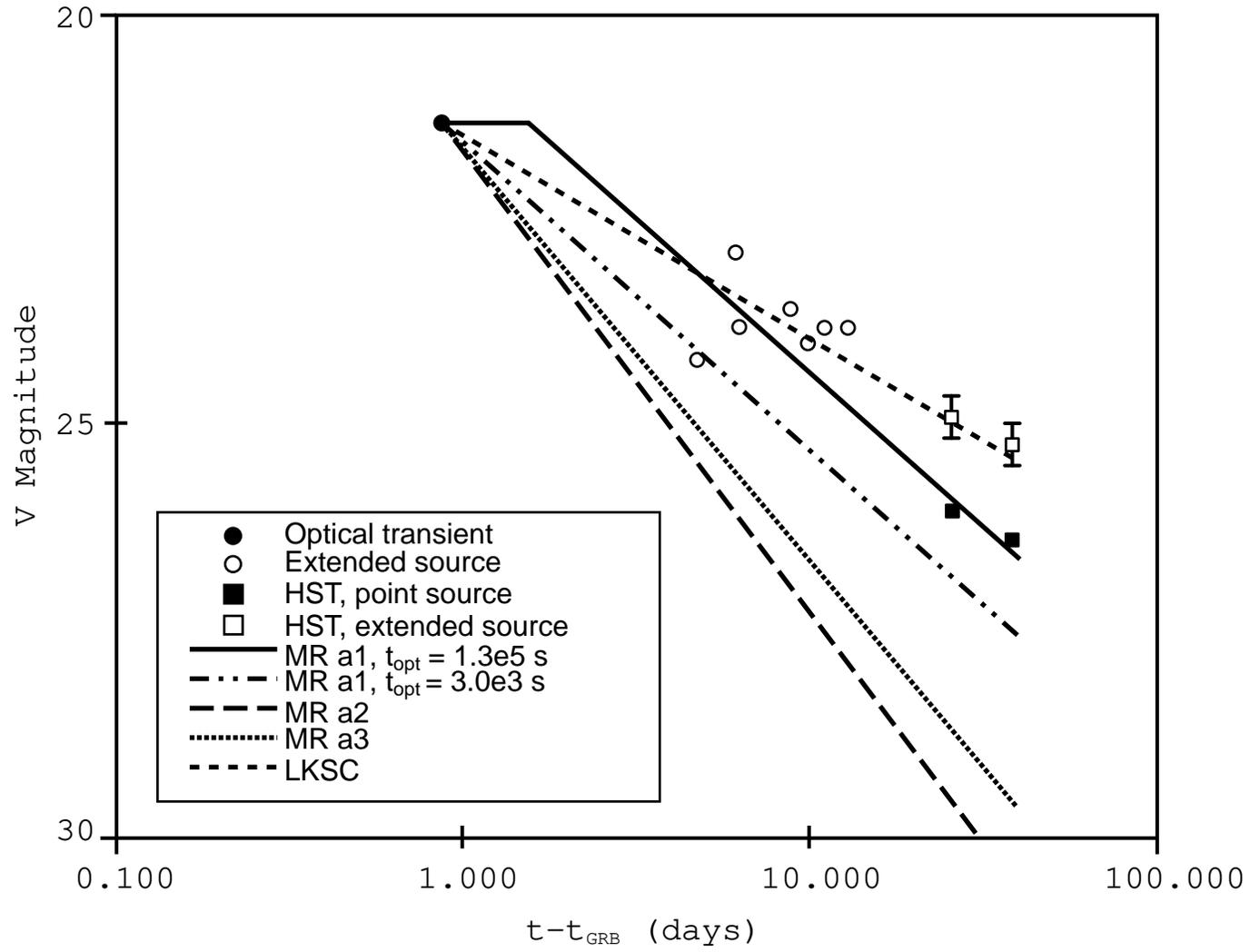
\end{document}